\documentclass[prc,twocolumn,showpacs]{revtex4}

\usepackage{graphicx}
\usepackage{dcolumn}
\usepackage{bm}
\usepackage{enumerate}

\usepackage{amsmath}
\usepackage{amssymb}
\usepackage{Ulem}
\usepackage{times}

\newcommand{\what}[1]{\widehat{#1}}

\newcommand{\bra}{\langle}
\newcommand{\ket}{\rangle}

\newcommand{\beq}{\begin{equation}}
\newcommand{\eeq}{\end{equation}}
\newcommand{\bea}{\begin{eqnarray}}
\newcommand{\eea}{\end{eqnarray}}

\def\la{\mathrel{\mathpalette\fun <}}

\def\fun#1#2{\lower3.6pt\vbox{\baselineskip0pt\lineskip.9pt
 \ialign{$\mathsurround=0pt#1\hfil##\hfil$\crcr#2\crcr\sim\crcr}}}


\begin{document}

\title{
Low lying excited states of $^{24}$O investigated by
self-consistent microscopic description \\ of proton inelastic scattering}

\author{Kazuhito Mizuyama}
\email{mizukazu147@gmail.com}
\author{Kazuyuki Ogata}

\affiliation{
Research Center for Nuclear Physics, Osaka University, Ibaraki 567-0047, Japan
}

\date{\today}

\begin{abstract}
The proton inelastic scattering of $^{24}$O($p,p'$) at 62 MeV/nucleon
is described by a self-consistent microscopic calculation with
the continuum particle-vibration coupling (cPVC) method.
The SLy5, SkM*, and SGII parameters are adopted as an effective
nucleon-nucleon interaction.
For all the parameters, the cPVC calculation reproduces very well the
first peak at 4.65~MeV in the $^{24}$O excitation energy spectrum
as well as its angular distribution.
The role of the cPVC self-energy strongly depends on the effective
interactions.
The higher-lying strength around 7.3~MeV is suggested to be
a superposition of the $3^-$ and $4^+$ states by the results
with SLy5 and SGII, whereas the SkM* calculation indicates
it is a pure $3^-$ state. This difference gives
a rather strong interaction dependence of
the angular distribution corresponding to the higher-lying strength.
\end{abstract}

\maketitle

Elucidation of particle-unbound excited states of neutron-dripline
nuclei is a hot subject in nuclear physics. It provides us with
rich information on the dripline nuclei that is difficult to extract
from their ground state properties.
Very recently, proton inelastic cross sections of $^{24}$O at 62~MeV/nucleon
were measured~\cite{24Oexp}.
The spin parity $\lambda^\pi$ of the first excited state at
$ 4.65\pm 0.14$~MeV was assigned $2^+$ and the quadrupole transition
parameter $\beta_2$ turned out to be small ($\beta_2 = 0.15 \pm 0.04$).
The large shell gap at $N=16$ ($N$ is the neutron number) was thus
confirmed. Another interesting finding of the measurement was
the strong (relatively-) higher-lying strength around 7.3~MeV considered to
have negative-parity configurations. This strength was suggestive
of the quenching of the gap between the $sd$ and $fp$ shells of neutron.
For more detailed discussion, clarification of the $\lambda^\pi$ of
the higher-lying strength will be necessary. Fully microscopic
description of the proton inelastic cross sections of $^{24}$O
is crucial for this purpose.

In a recent paper~\cite{CPVC}, the continuum particle-vibration
coupling (cPVC) method was proposed and applied to studies on the
single-particle (sp) structures in $^{40}$Ca, $^{208}$Pb, and $^{24}$O.
It was shown in Ref.~\cite{CPVC} that the cPVC
method describes quite well the fragmentation of the sp hole
and particle states as well as the shift of those centroid
energies, in good agreement with the experimental spectroscopic
factor.

We then applied the cPVC method to the neutron elastic scattering
on $^{16}$O at 4--30~MeV and succeeded in reproducing experimental data of
the total-elastic cross section $\sigma_{\rm el}$ and
the reaction cross section $\sigma_{\rm R}$ satisfactorily well~\cite{ELCPVC}.
The cPVC method treats the neutron single-particle motion and
the vibration of the $^{16}$O nucleus in a unified manner.
One of the remarkable aspects of the cPVC method for the $n+{}^{16}$O
system is that the important properties of the neutron optical potential,
i.e., energy dependence, complex nature, and non-locality, are automatically
generated within a coupled-channel framework using an energy-independent
and real nucleon-nucleon effective interaction. This gives a fully
microscopic self-consistent description of nucleon-nucleus scattering.
Furthermore, the cPVC method can describe even doorway states~\cite{Wei61} of the
nucleon-nucleus system that are observed as narrow peaks in the
total reaction cross section. It is well known that such states
cannot be described by a phenomenological optical potentials or by
a folding model calculation based on a complex $g$ matrix
~\cite{Deb05,Ray92,Coo93,Rik84,RG84,Amo00,DA00,Chi98,Fur08}.

In the description of neutron elastic scattering by the cPVC method,
large numbers of discrete and continuum phonon states of the core nucleus
(or target in a terminology of reaction studies)
are explicitly taken into account. Therefore,
it is quite straightforward to extend the cPVC method to inelastic
scattering processes. The purpose of the present work is to
apply the cPVC method to the proton inelastic scattering by $^{24}$O
and to discuss the spin-parity of the higher-lying strength around 7.3~MeV
observed.

\begin{figure}[htpb]
\begin{center}
\includegraphics[width=0.45\textwidth,clip]{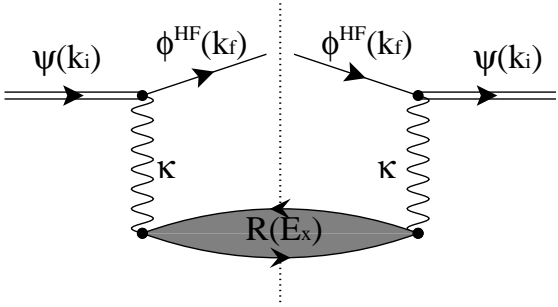}
\caption{
The Feynman diagram corresponding to the double differential cross section
for the nucleon-nucleus inelastic scattering expressed by Eq.~(\ref{sig2}).
$\phi^{\rm HF}$ and $\psi$ are the HF and cPVC scattering wave functions,
respectively.
$\kappa$ is the residual force. $R(E_x)$ is the continuum RPA response function.
The half part of the diagram expresses the transition matrix.
}
\label{fig1}
\end{center}
\end{figure}

The double differential cross section for the proton-nucleus inelastic
scattering is given by
\begin{eqnarray}
\frac{d^2\sigma}{dE_xd\Omega}
&=&
-
\left(\frac{4 m}{\hbar^2}\right)^2
\frac{k_f}{k_i}
\sum_\lambda
\mbox{ Im }
\sum_{l_f l_i}
\sum_{l_f' l_i'}
{\cal P}_{\lambda}^{l_fl_i;l_f'l_i'}(\cos\theta)
\nonumber\\
&&\hspace{-1.2cm}
\times
\int \!\!dr\!\!
\int \!\!dr'
\bra C\kappa_{l_fl_i}^{\lambda}(r)\ket^*
R_{\lambda}(r,r';E_x)
\bra C\kappa_{l_f'l_i'}^{\lambda}(r')\ket,
\label{sig2}
\end{eqnarray}
where
\begin{equation}
\bra C\kappa_{l_fl_i}^{\lambda}(r)\ket
\equiv
\sum_{j_fj_i}
C_{l_fj_f;l_ij_i}^{\lambda}
\bra \phi^{\rm HF}_{l_fj_f}(k_f)||\kappa(r)||\psi_{l_ij_i}(k_i)\ket,
\end{equation}
\begin{eqnarray}
C_{l_fj_f;l_ij_i}^{\lambda}
&\equiv&
\frac{(-)^{j_i+l_f+\frac{1}{2}+\lambda}}{\sqrt{2}}
\frac{
\what{j}_f
\what{j}_i}
{\what{\lambda}^3}
\nonumber\\
&&\times
\bra l_f j_f ||Y_{\lambda}||l_i j_i\ket
\left\{
\begin{array}{ccc}
l_i & j_i & 1/2 \\
j_f & l_f & \lambda
\end{array}
\right\},
\end{eqnarray}
\begin{eqnarray}
&&\hspace{-1.2cm}
\bra \phi^{\rm HF}_{l_fj_f}(k_f)||\kappa(r)||\psi_{l_ij_i}(k_i)\ket
\nonumber\\
&\equiv&
i^{l_i}(i^{l_f})^*
\phi^*_{HF,l_fj_f}(r;k_f)
\frac{\kappa(r)}{r^2}
\psi_{l_i j_i}(r;k_i),
\end{eqnarray}
and
\begin{eqnarray}
&&\hspace{-0.6cm}
{\cal P}_{\lambda}^{l_fl_i;l_f'l_i'}(\cos\theta)
\nonumber\\
&\equiv&
\sum_I
\what{\lambda}^2
\left\{
\begin{array}{ccc}
l_i & l_f & \lambda \\
l_f' & l_i' & I
\end{array}
\right\}
\bra l_i'||Y_I||l_i\ket\bra l_f'||Y_I||l_f\ket
P_I(\cos\theta).
\nonumber\\
\end{eqnarray}
The excitation energy of the target nucleus is denoted by $E_x = E_i-E_f$
with $E_c=\hbar^2 k_c^2/(2m)$; $m$ is the nucleon mass and $k_c$ is the wave
number of proton. $c=i$ and $f$ represent the initial and final states,
respectively. The orbital (total) angular momentum of proton is denoted by
$l_c$ ($j_c$).
$\psi_{l_ij_i}$ is the cPVC scattering wave function obtained by solving
the cPVC Lippmann-Schwinger equation~\cite{ELCPVC}. We use $\psi_{l_ij_i}$ in the
transition matrix as the total wave function of the system in the initial state.
$\phi^{\rm HF}_{l_fj_f}(k_f)$ is the Hartree-Fock (HF) scattering wave function,
$R_{\lambda}$ is the response function calculated by the continuum random phase approximation
(RPA)~\cite{mizuyama,sagawa}, and $\kappa(r)$ is the residual interaction. For simplicity, we
disregard the spin-dependent terms of the residual interaction in the calculation
of RPA and cPVC self-energy.
The momentum-dependent terms are taken into account self-consistently in RPA,
but the Landau-Migdal (LM) approximation~\cite{CPVC} is adopted in the
calculation of the self-energy.
 Furthermore, the residual Coulomb interaction
is dropped in the self-energy function.
The Feynman diagram corresponding to Eq.~(\ref{sig2}) is shown in Fig.~\ref{fig1}.

%
In the present study, the RPA phonons
of $\lambda^\pi=1^-,2^+,3^-,4^+$, and $5^-$ are included and the maximum
excitation energy of $^{24}$O is taken to be 60~MeV.
The angular momentum cut-off $l_{\rm max}$ for the single-particle
 orbits is set to 12.  As the nucleon-nucleon effective interaction,
we take SLy5~\cite{SLy5}, SkM*~\cite{SkMs}, and SGII\cite{SGII} Skyrme parameters,
and discuss the interaction dependence of the inelastic cross section.
We choose the Fermi momentum $k_{\rm F}=1.33$~fm$^{-1}$ for
the residual force with the LM approximation.
 The radial mesh size is $0.2$~fm and the maximum value of
$r$ in the integral in Eq.~(\ref{sig2}) is set to $20$~fm.

\begin{figure}[htpb]
\begin{center}
\includegraphics[width=0.5\textwidth,clip]{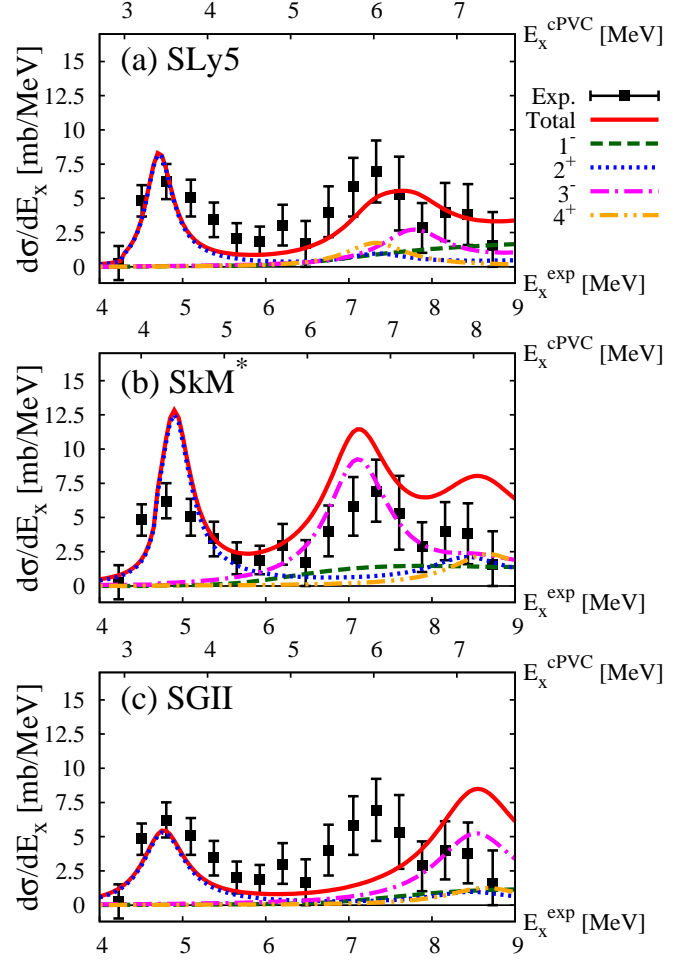}
\caption{(Color online)
The differential cross section $d\sigma/dE_x$ plotted as a function
of the excitation energy $E_x$ of $^{24}$O for the $(p,p')$ reaction at
62~MeV/nucleon, and compared with the experimental data~\cite{24Oexp}.
Panels (a), (b), and (c) are the results with the SLy5, SkM*, and SGII
Skyrme interaction, respectively. The solid line is the total energy spectrum.
The dashed, dotted, dash-dotted and dash-dot-dashed lines are
the contributions from the $1^-, 2^+, 3^-$,
and $4^+$ states of $^{24}$O, respectively, in each panel.
The lower horizontal axis is the experimental excitation energies of $^{24}$O,
$E_x^{\rm exp}$, and the upper axis is the theoretical one, $E_x^{\rm cPVC}$.
See the text for detail.
}
\label{fig2}
\end{center}
\end{figure}
In Fig.~\ref{fig2} we show the energy spectra of the
$^{24}$O($p,p'$) reaction at 62~MeV calculated by the cPVC method
with the (a) SLy5, (b) SkM*, and (c) SGII interactions.
In each panel, the solid line is the total energy spectrum
plotted as a function of $E_x$ and compared with the experimental data;
the dashed, dotted, dash-dotted,
and dash-dot-dotted lines respectively show the contributions from
the $1^-$, $2^+$, $3^-$, and $4^+$ states of $^{24}$O. We have smeared
the theoretical results taking account of the experimental resolution
$0.5\sqrt{E_x - 4.09}$ in FWHM given in Ref.~\cite{24Oexp}.
Additionally, we have shifted the theoretical cross sections
obtained with the SLy5 and SGII interactions
by $\Delta E_x =1.3$~MeV toward the high $E_x$ direction,
to adjust its
first peak to that of the experimental data;
for the result with SkM*, $\Delta E_x =0.5$~MeV is used.
We have not renormalized the absolute values of the cross sections
shown in this paper.
One sees from Fig.~\ref{fig2} that for SLy5 and SkM* the cross
sections calculated agree well with the experimental data, which
shows the success of the
cPVC method in describing the proton inelastic scattering.
When SGII is used, however, the relative energy between the two peaks
is not reproduced well.
For all the interactions,
the first peak around 4.65~MeV is found to be purely the $2^+$ state of
$^{24}$O, as confirmed in Ref.~\cite{24Oexp}.

There seems to be a missing strength for 5~MeV~$\la E_x \la$~6~MeV for
the results with the SLy5 and SGII interactions. This strength
may be due to contributions from unnatural parity states,
the $1^+$ state in particular, which are not included in the present
calculation. It was claimed in Ref.~\cite{24Oexp} that
the $1^+$ contribution was too small to explain the strength
around 5~MeV~$\la E_x \la$~6~MeV,
which is suggested by also an RPA calculation~\cite{Ina13},
and it might be due to the contribution of negative parity states.
The dashed and dash-dotted lines
in Figs.~\ref{fig2}(a) and \ref{fig2}(c)
indicate that there is no contribution of the
negative parity states in this region.
On the other hand, we have no such missing strength when the SkM*
interaction is used. In panel (b) the peak around 4.65~MeV
is due to almost the $2^+$ state, with a small background of
the $3^-$ state. Thus, the result for 5~MeV~$\la E_x \la$~6~MeV
shows the effective interaction dependence.

The interaction dependence is more significant for the
higher-lying strength around 7.3~MeV. If SLy5 is used it is a superposition
of the $3^-$ and $4^+$ strengths added by the $1^-$ and $2^+$
background contributions. On the other hand, the strength almost comes
from the $3^-$ state when SkM* is adopted.
The result with the SGII interaction is qualitatively the same as
that with SkM*. However, the relative energy between the two peaks
is not properly obtained as mentioned above.
It should be noted that Skyrme parameters are determined to reproduce
the properties of the ground state and giant resonances of nuclei
in middle- and heavy-mass regions. On the other hand, it is rather well
known that the low-lying
excited states are more sensitive to the shell structure and may not
necessarily be described well by the Skyrme interactions.
Therefore, it is not surprising
that the reaction observables associated with the low-lying states of
the target nucleus show rather strong dependence on
the Skyrme parameters.
Considering this, the success of the cPVC method with Skyrme interactions
in reproducing the experimental data will be quite remarkable, though
there remain some discrepancies between the calculations and the experimental
data.
It will be interesting and important
to use reaction observables as new constraints on the Skyrme parameters.

\begin{figure}[htpb]
\begin{center}
\includegraphics[width=0.5\textwidth,angle=-90,clip]{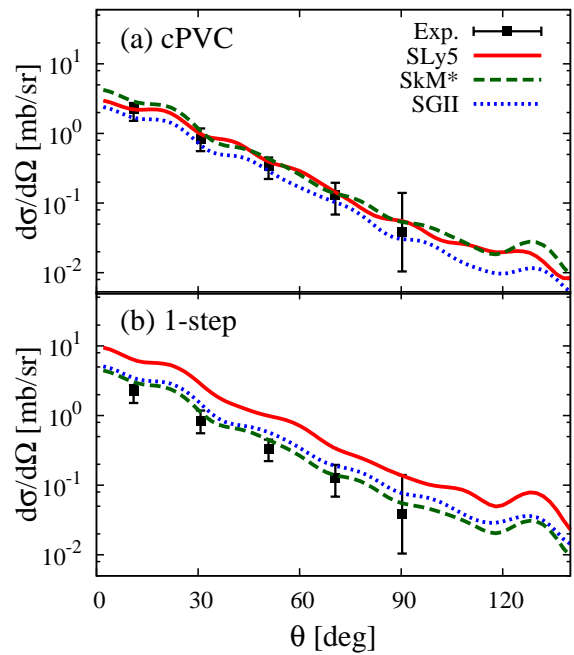}
\caption{(Color online) The angular distribution of the cross section $d\sigma/d\Omega$
corresponding to the first $2^+$ state of $^{24}$O at $E_x=4.65\pm0.14$~MeV compared with
the experimental data~\cite{24Oexp}.
The solid, dashed, and dotted lines are the results with the SLy5, SkM*, and SGII
respectively. Panels (a) and (b) are the results by the cPVC and a one-step transition respectively~(see text).
}
\label{fig3}
\end{center}
\end{figure}
Figure~\ref{fig3}(a) shows the angular distribution corresponding to
the first $2^+$ peak at $E_x=4.65\pm0.14$~MeV. The results
with SLy5, SkM*, and SGII are plotted by the solid, dashed, and dotted lines,
respectively. One sees all the calculations perfectly reproduce the
experimental data; the interaction dependence seems very small.
However, if we see the coupled-channel effects, a clear
interaction dependence appears. The results shown in Fig.~\ref{fig3}(b)
correspond to a one-step transition from the HF ground
state of $^{24}$O to its RPA $2^+$ state; the motion of
the incoming (outgoing) proton is described by a mean-field
potential generated by the ground (RPA $2^+$) state of $^{24}$O.
In other words, only the folding part of an optical potential
is included dropping all the dynamical polarization potentials (DPPs).
Note that this one-step calculation is different from
a usual DWBA calculation that adopts the proton-target distorting
(optical) potentials including DPPs in the initial and final channels.
The solid line in Fig.~\ref{fig3}(b) is significantly larger than
that in Fig.~\ref{fig3}(a),
mainly because of the absence of absorption, i.e., the imaginary
part of the optical potential. A similar difference is found in
the results with the SGII interaction.
On the other hand,
the one-step calculation with SkM* is almost
the same as the result of the cPVC calculation.
This indicates that there is no absorption of the flux of the
incident proton when SkM* is adopted, which is quite difficult
to understand. If this is indeed the case, the proton elastic
scattering on $^{24}$O can be described by only the HF single-particle
potential, suggesting a very small value of reaction cross section.

\begin{figure}[htpb]
\begin{center}
\includegraphics[width=0.58\textwidth,angle=-90,clip]{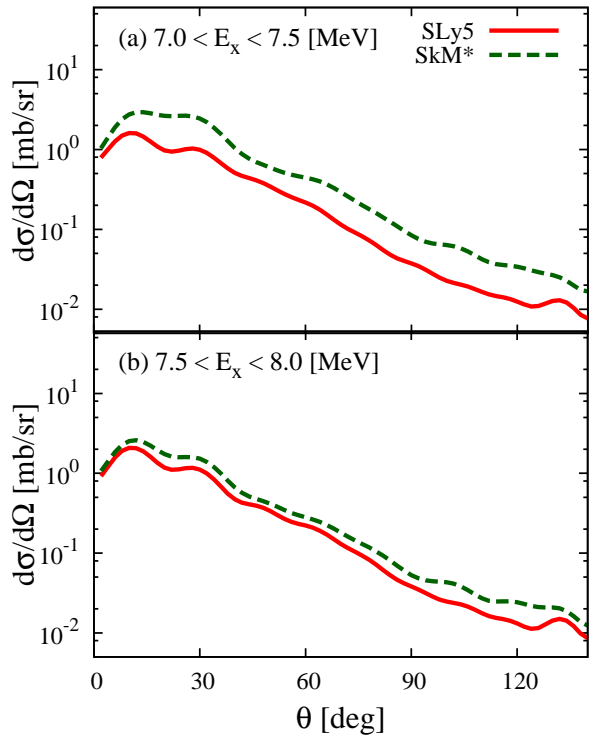}
\caption{(Color online) The angular distributions correspond to
(a) 7~MeV~$\le E_x \le$~7.5~MeV and
(b) 7.5~MeV~$\le E_x \le$~8~MeV. In each panel the solid
and dashed lines show the results with the SLy5 and SkM* interactions,
respectively.}
\label{fig4}
\end{center}
\end{figure}
The effective-interaction dependence of the result
will be investigated more clearly
if the angular distribution of the higher-lying states is obtained.
Figure~\ref{fig4} shows the theoretical results of the angular
distribution; panels (a) and (b) correspond to the energy
regions of 7~MeV~$\le E_x \le$~7.5~MeV and 7.5~MeV$\le E_x \le$~8~MeV,
respectively. The solid (dashed) line in each panel shows the
result calculated with the SLy5 (SkM*) interaction. The shape
of the solid line in panel (a) is quite different from that
in panel (b). This is because, as shown in Fig.~\ref{fig2}(a),
the spin-parity of $^{24}$O having the largest contribution
is $4^+$ for 7~MeV~$\le E_x \le$~7.5~MeV and $3^-$ for
7.5~MeV~$\le E_x \le$~8~MeV when SLy5 is adopted. On the other hand,
the results with SkM* (the dashed lines) have a similar shape,
reflecting the fact that the $3^-$ state has a dominant contribution
in both energy ranges, as shown in Fig.~\ref{fig2}(b).
%
\begin{figure}[htpb]
\begin{center}
\includegraphics[width=0.6\textwidth,angle=-90,clip]{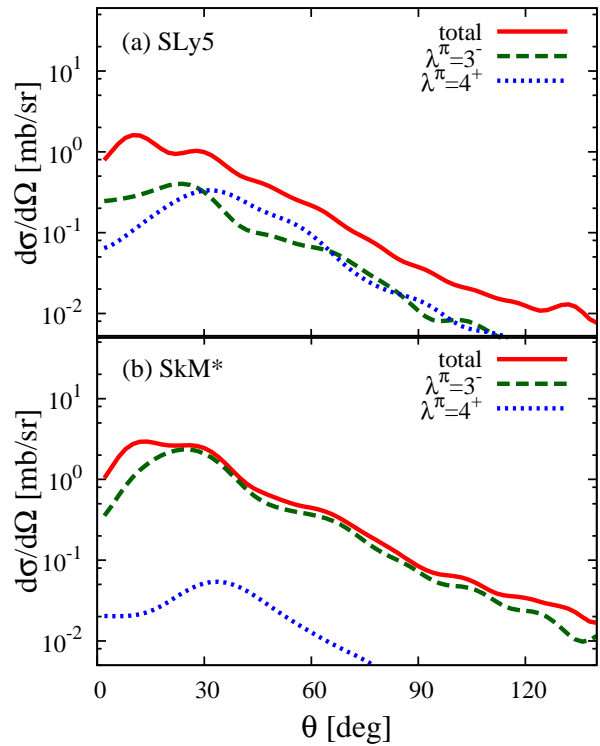}
\caption{(Color online) The contribution of the $3^-$(dashed line) and
$4^-$(dotted line) states to the angular distribution
for $7.0~\le~E_x~\le~7.5$~MeV shown by Fig.~\ref{fig4}(a).
Pannels (a) and (b) correspond to the results
with SLy5 and SkM*, respectively. }
\label{fig5}
\end{center}
\end{figure}
In Fig.~\ref{fig5} we decompose the results in Fig.~\ref{fig4}(a)
into the contributions of the $3^-$ and $4^+$ states.
Clearly, the difference in the $4^+$ state contributions for SLy5 and
SkM* gives the different angular distributions for
7~MeV~$\le E_x \le$~7.5~MeV.
Comparison with experimental data will be very interesting and important.

In summary, we have applied the cPVC method to the $^{24}$O($p,p'$) reaction
at 62 MeV/nucleon. We took the SLy5, SkM*, and SGII parameters
as an effective nucleon-nucleon interaction. For all the three parameters,
the cPVC calculation well reproduces the experimental data of
the energy spectrum and the angular distribution corresponding to
the first $2^+$ state, except that the relative energy between the
two peaks cannot be reproduced when SGII is adopted.
One finds that the calculations underestimate the strength
at 5~MeV~$\la E_x \la$~6~MeV when SLy5 or SGII is adopted, which may imply
some contributions from unnatural parity states.
On the other hand, there is no missing strength in that region
when SkM* is used.
The role of the cPVC self-energy, i.e., the dynamical polarization
potential, has a strong dependence on the effective interactions.
The higher-lying strength around 7.3~MeV is found to be
a superposition of the $3^-$ and $4^+$ states by the calculation
with SLy5 and SGII, whereas the SkM* calculation indicates
the strength is almost due to the $3^-$ state.
Reflecting this difference, the shape of the angular distribution
of the higher-lying strength has a rather strong dependence on the
effective interaction. Experimental data for this distribution
may judge which interaction is preferable for describing
the proton inelastic scattering.

The authors thank G.~Col\`o, E.~Vigezzi, T.~Inakura, H.~Nakada, and
M.~Yahiro for helpful discussions.
This research was supported in part by Grant-in-Aid of the Japan
Society for the Promotion of Science (JSPS).


\begin{thebibliography}{00}
\bibitem{24Oexp}
K.~Tshoo, {\it et.al.}, Phys.~Rev.~Lett {\bf 109}, 022501(2012).

\bibitem{CPVC}
K.~Mizuyama, G.~Col\`o, and E.~Vigezzi,
Phys. Rev. C {\bf 86}, 034318 (2012).

\bibitem{ELCPVC}
K.~Mizuyama and K.~Ogata,
Phys. Rev. C {\bf 86}, 041603(R)(2012).

\bibitem{Wei61}
C.~F.~Weisskopf,
Phys. Today {\bf 14}, 18 (1961).

\bibitem{Deb05}
P.~K.~Deb, B.~C.~Clark, S.~Hama, K.~Amos, S.~Karataglidis,
and E.~D.~Cooper,
Phys. Rev. C {\bf 72}, 014608 (2005).

\bibitem{Ray92}
L.~Ray, G.~W.~Hoffmann, and W.~R.~Coker,
Phys. Rep. {\bf 212}, 223 (1992).

\bibitem{Coo93}
E.~D.~Cooper, S.~Hama, B.~C.~Clark, and R.~L.~Mercer,
Phys. Rev. C {\bf 47}, 297 (1993).

\bibitem{Rik84}
L. Rikus, K. Nakano and H. V. von Geramb,
Nucl. Phys. {\bf A414}, 413 (1984).

\bibitem{RG84}
L. Rikus and H.V. von Geramb, Nucl. Phys. {\bf A426}, 496 (1984).

\bibitem{Amo00}
K. Amos, P. J. Dortmans, H. V. von Geramb, S. Karataglidis,
and J. Raynal,
Adv. Nicl. Phys. {\bf 25}, 275 (2000).

\bibitem{DA00}
P. K. Deb and K. Amos, Phys. Rev. C {\bf 62}, 024605 (2000).

\bibitem{Chi98}
S. P. Weppner, Ch. Elster, and D. Huber,
Phys. Rev. C {\bf 57}, 1378 (1998) and references therein.

\bibitem{Fur08}
T. Furumoto, Y. Sakuragi, and Y. Yamamoto, Phys. Rev. C {\bf 78},
044610 (2008),
{\it ibid.}, C {\bf 79}, 011601(R) (2009),
{\it ibid.}, C {\bf 80}, 044614 (2009).

\bibitem{mizuyama}
K.~Mizuyama, M.~Matsuo, and Y.~Serizawa,
Phys. Rev. C {\bf 79}, 024313 (2009).

\bibitem{sagawa}
H.~Sagawa, Prog. Theor. Phys. Suppl. {\bf 142}, 1 (2001).

\bibitem{SLy5}
E.~Chabanat, P.~Bonche, P.~Haensel, J.~Meyer and R.~Schaeffer,
Nucl. Phys. {\bf A635}, 231 (1998).

\bibitem{SkMs}
J. ~Bartel, P. ~Quentin, M.~Brack, C.~Guet and H.~-B.~H\r{a}kansson,
Nucl. Phys. {\bf A386}, 79 (1982).

\bibitem{SGII}
Nguyen ~Van ~Giai and H.~Sagawa, Nucl.~Phys.~{\bf A371}, 1(1981).

\bibitem{Ina13}
T. Inakura, private communication (2013).

\end{thebibliography}
\end{document}